\documentclass[preprint,eqsecnum,aps,nofootinbib]{revtex4}
\usepackage[dvips]{graphicx}
\usepackage{amssymb}
\usepackage{amsmath}
\usepackage{amsmath, amsthm, amssymb, mathrsfs}

\begin{document}

\title{Bulk versus Brane Emissivities of Photon Fields:  
For the case of Higher-Dimensional Schwarzschild Phase} 
\author{Eylee Jung\footnote{Email:eylee@kyungnam.ac.kr}}
\author{D. K. Park\footnote{Email:dkpark@hep.kyungnam.ac.kr}}

\affiliation{Department of Physics, Kyungnam University, Masan, 631-701, Korea}

\begin{abstract}
The emission spectra for the spin-$1$ photon fields are computed when the 
spacetime is a $(4+n)$-dimensional Schwarzschild phase. For the case of the 
bulk emission we compute the spectra for the vector mode and scalar mode
separately. Although the emissivities for the scalar mode is larger than 
those for the vector mode when $n$ is small, the emissivities for the 
vector mode photon become dominant rapidly with increasing $n$. For the 
case of the brane emission the emission spectra are numerically computed
by making use of the complex potential method. Comparision of the total bulk
emissivities with total brane emissivities indicates that the 
effect of the field
spin makes the bulk emission to be rapidly dominant with increasing $n$.
However, the bulk-to-brane relative emissivity per degree of freedom always 
remains smaller than unity.
The importance for the spin-$2$ graviton emission problem is discussed. 
\end{abstract}


\maketitle

\newpage
\section{Introduction}
The assumption for the existence of the extra dimensions has a 
long history\cite{hclee84,ruba83}. Recent quantum gravity such as 
string theories\cite{polchin98} and brane-world scenarios\cite{bwsc1}
also adopts this assumption. Especially modern brane-world scenarios
predict the emergence of the TeV-scale gravity, which opens the possibility
to make tiny black holes by high-energy scattering in the future
collider\cite{hec1}. In this reason much attention is paid recently to the
higher-dimensional black holes\cite{tang63}.

The most well-known quantum gravity effect of the black hole is a 
Hawking radiation\cite{hawking75}, which makes the black hole different from
the black body. The Hawking radiation for the higher-dimensional black holes
was extensively studied for last few years to support the experimental 
significance in the future colliders. Emparan, Horowitz and Myers (EHM)
argued that the brane-world black holes radiate mainly on the 
brane\cite{emp00}. To support their argument roughly EHM used a simple setting
like an higher-dimensional black body.

More exact calculation on the absorption and emission problems for the 
higher-dimensional non-rotating black holes was performed in
Ref.\cite{kanti1,jung05-1}. The authors of Ref.\cite{kanti1} carried out the
numerical calculation in the background of the $(4+n)$-dimensional 
Schwarzschild black hole. In Ref.\cite{jung05-1} different numerical technique
was adopted and the Hawking radiation by the $(4+n)$-dimensional charged 
black hole was explored. The numerical results of Ref.\cite{kanti1,jung05-1}
support the EHM argument, {\it black holes radiate mainly on the brane}, if 
$n$ is not too large.

More recently, there was a suggestion that EHM argument should be examined 
carefully when the black holes have angular momenta\cite{frol1}. When the 
fields are scattering with the rotating black hole, there is a factor called
superradiance\cite{superr1}, which does not exist in the case of the 
non-rotating black hole. The authors in Ref.\cite{frol1} argued that EHM
argument may be wrong due to the existence of the superradiance modes. The
condition for the existence of the superradiance modes was derived for the
bulk fields\cite{jung05-2} and brane-localized fields\cite{ida02}.
In Ref.\cite{jung05-3} numerical calculation was performed for the bulk 
and brane-localized scalar fields in the background of the five-dimensional
rotating black holes with two different angular momenta. According to 
Ref.\cite{jung05-3} the numerical calculation shows that the energy 
amplification for the bulk scalar is very small (roughly order of 
$10^{-9} \%$) while that for the brane scalar is order of unity. This big
difference indicates that the effect of the superradiance is negligible
for the case of the bulk scalar. Therefore, the standard claim, 
{\it black holes radiate mainly on the brane}, still holds although 
the effect of the superradiance is taken into account. 

There is an another factor we should check carefully when the Hawking radiation
for the higher-dimensional black holes is studied. This is an effect for the 
higher-spin particles like graviton. Since the graviton is not generally
localized on the brane unlike the usual standard model particles, the 
EHM argument should be carefully re-checked in the graviton emission.
When the spacetime background is a $(4+n)$-dimensional Schwarzschild metric
\begin{equation}
\label{metric1}
ds^2 = - h dt^2 + h^{-1} dr^2 + r^2 d \Omega_{n+2}^2
\end{equation}
where $h = 1 - (r_H / r)^{n+1}$ and  the angular part $d \Omega_{n+2}^2$ is a 
spherically symmetric line element in a form
\begin{equation}
\label{angle-part}
d\Omega_{n+2}^2 = d\theta_{n+1}^2 + 
\sin^2 \theta_{n+1} \Bigg[ d\theta_{n}^2 + \sin^2 \theta_n \bigg(
\cdots + \sin^2 \theta_2 \left( d\theta_1^2 + \sin^2 \theta_1 d\varphi^2
           \right) \cdots \bigg) \Bigg],
\end{equation}
the radial equations for the bulk graviton were derived in Ref.\cite{koda03}
by extending the well-known Regge-Wheeler method\cite{reg57}. Using these
radial equations, the absorption and emission spectra for the scalar,
vector, and tensor modes of the bulk graviton were computed 
recently\cite{corn95}. It was shown that the total emissivity for the 
bulk graviton increases rapidly compared to the
spin-$0$ bulk scalar when the extra dimensions exist. The ratio of the total
emissivities between the bulk graviton and bulk scalar, for example, becomes
$5.16 \%$, $147.7 \%$, $595.2 \%$ and $3417 \%$ when $n=0$, $1$, $2$ and
$6$ respectively. This tremendous increase of the emission rate for the 
bulk graviton makes us believe that the missing energy is not negligible 
in the future collider experiment although it is not larger than the visible
one.  
In order to compare the graviton emissivities on the brane and in the bulk we 
should 
compute the emission rate for the brane-localized graviton. In order to 
explore the absorption and emission problems for the brane graviton
the axial and polar perturbations were studied in Ref.\cite{park06-1} when 
the spacetime background is a $4d$ induced metric from Eq.(\ref{metric1})
\begin{equation}
\label{induced}
ds_4^2 = -h dt^2 + h^{-1} dr^2 + r^2 (d \theta^2 + \sin^2 \theta d\phi^2).
\end{equation}
In these perturbations there is a difficulty arising due to the fact that
the metric (\ref{induced}) is not a vacuum solution of the $4d$ Einstein
field equation. Thus in Ref.\cite{park06-1} the metric (\ref{induced}) was 
regarded as the non-vacuum solution. This yields an additional 
difficulty on the 
treatment of the energy-momentum tensor in the given perturbation.

In order to explore the effect of the field spin in the absorption and 
emission problems we choose the spin-$1$ photon field in the background of the 
$(4+n)$-dimensional Schwarzschild phase (\ref{metric1}). 
The brane emission rates for the spin-$1$ field were already computed in the 
background of non-rotating black hole in Ref.\cite{kanti1} and rotating 
black hole in the last reference of Ref.\cite{ida02}. In this paper we will
compute the bulk emission rate too to analyze the validity of the EHM argument
in the Hawking evaporation of the spin-$1$ fields.
In Sec. II we compute
the absorption and emission spectra for the bulk photon fields. For the 
comparision we compute the spectra for the vector mode photon and scalar
mode photon separately. In Sec. III we compute the emission spectra for the 
brane-localized photon fields by making use of the induced 
metric (\ref{induced}). In the next section the emission rates for the 
bulk photon fields are compared to those for the brane-localized photon
fields. It is shown that the brane emission is a bit dominant when $n \leq 4$.
However, the dominance is changed into the total bulk emission when $n \geq 6$.
However the bulk-to-brane emissivity per degree of freedom always remains 
smaller than unity, which supports the EHM argument. 
In Sec. V a brief conclusion is given.

\section{Bulk Photon}
The electromagnetic perturbation in the background of the higher-dimensional
spherically symmetric black holes was discussed in detail in Ref.\cite{cris01}.
The most remarkable fact is that the perturbations are classified into the 
scalar- and vector-type modes according to their tensorial behavior on the 
spherical section of the background metric. 
The radial equations for the modes are expressed as a Schr\"{o}dinger-like
equation in the form 
\begin{equation}
\label{bl-radial1}
\left[ \left( h \frac{d}{d r} \right)^2 + \omega^2 \right] R = V_{BL} R
\end{equation}
where the effective potential $V_{BL}$ is 
\begin{equation}
\label{eff-p1}
V_{BL} = \frac{h}{r^2} \left[\ell (\ell + n + 1) + 
         \frac{n (n + 2)}{4} + \sigma_n (1 - h) \right]
\end{equation}
with $\ell \geq 1$.
The $n$-dependent parameter $\sigma_n$ is also dependent on the modes 
for the bulk photon in the following
\begin{eqnarray}
\label{eff-p2}
\sigma_n = \left\{ \begin{array}{ll}
            \frac{n (n + 4)}{4}   &   \mbox{for vector photon mode}   \\
            -\frac{n (3 n + 4)}{4}  &  \mbox{for scalar photon mode}  \\
            \frac{(n + 2)^2}{4}   &   \mbox{for bulk scalar field}
                  \end{array}
                  \right.
\end{eqnarray}
Defining the dimensionless parameters $x \equiv \omega r$ and 
$x_H \equiv \omega r_H$, one can rewrite Eq.(\ref{bl-radial1}) as following:
\begin{eqnarray}
\label{bl-radial2}
& &
\hspace{2.0cm} 
x^2 (x^{n+1} - x_H^{n+1})^2 \frac{d^2 R}{d x^2} + (n + 1) x_H^{n + 1}
x (x^{n+1} - x_H^{n+1}) \frac{dR}{d x}
                                               \\   \nonumber
& & + \left[x^{2 n + 4} - (x^{n+1} - x_H^{n+1})
            \left\{ \left[ \ell (\ell + n + 1) + \frac{n (n + 2)}{4} \right]
            x^{n + 1} + \sigma_n x_H^{n + 1} \right\}   \right] R = 0.
\end{eqnarray}
Since Eq.(\ref{bl-radial2}) is real, it is easy to show that if $R$ is a 
solution of Eq.(\ref{bl-radial2}), its complex conjugate $R^*$ is solution too.
The Wronskian between them is 
\begin{equation}
\label{wron1}
W[R^*, R]_x \equiv R^* \frac{d R}{d x} - R \frac{d R^*}{d x}
= {\cal C} \frac{x^{n + 1}}{x^{n+1} - x_H^{n+1}}
\end{equation}
where ${\cal C}$ is an integration constant.

Now, we consider the solution of Eq.(\ref{bl-radial2}), 
${\cal G}_{n,\ell}^{BL}$, which is convergent in the neighborhood of the 
near-horizon $x \sim x_H$. Since $x = x_H$ is a regular singular point of 
the radial equation (\ref{bl-radial2}), we can express ${\cal G}_{n,\ell}^{BL}$
as a convergent series in the form:
\begin{equation}
\label{bl-near1}
{\cal G}_{n,\ell}^{BL} (x, x_H) = \sum_{N=0}^{\infty} d_{\ell,N} 
(x - x_H)^{N+\rho_n}
\end{equation}
where $\rho_n = -i x_H / (n + 1)$. Making use of Eq.(\ref{wron1}), it is easy
to derive the Wronskian between ${\cal G}_{n,\ell}^{BL *}$ and 
${\cal G}_{n,\ell}^{BL}$:
\begin{equation}
\label{wron2}
W[{\cal G}_{n,\ell}^{BL *}, {\cal G}_{n,\ell}^{BL}]_x = -2 i 
|g_{n,\ell}|^2 \frac{x^{n + 1}}{x^{n+1} - x_H^{n+1}}
\end{equation}
where $g_{\ell,n} \equiv d_{\ell,0}$. Inserting Eq.(\ref{bl-near1}) into
(\ref{bl-radial2}), one can directly derive the recursion relation for the 
coefficients $d_{\ell,N}$. The recursion relation is ,of course, $n$-dependent 
and lengthy. Therefore, we will not present it explicitly.

Next we consider the solutions of Eq.(\ref{bl-radial2}), 
${\cal F}_{n,\ell(\pm)}^{BL}$, which are convergent in the asymptotic regime:
\begin{equation}
\label{bl-asymp1}
{\cal F}_{n,\ell(\pm)}^{BL} (x, x_H) = (\pm i)^{\ell + 1 + n/2} x 
e^{\mp i x} (x - x_H)^{\pm \rho_n}
\sum_{N=0}^{\infty} \tau_{N(\pm)} x^{-(N+1)}
\end{equation}
with $\tau_{0(\pm)} = 1$. The solutions ${\cal F}_{(+)}$ and ${\cal F}_{(-)}$
are, therefore, the ingoing and outgoing waves respectively. The Wronskian
between them is 
\begin{equation}
\label{wron3}
W[{\cal F}_{n,\ell(+)}^{BL}, {\cal F}_{n,\ell(-)}^{BL}]_x = 2 i 
\frac{x^{n+1}}{x^{n+1} - x_H^{n+1}}.
\end{equation}
As in the previous case the recursion relation for $\tau_{N(\pm)}$ can be 
derived explicitly by inserting (\ref{bl-asymp1}) into (\ref{bl-radial2}).

We assume the real scattering solution ${\cal R}_{n,\ell}$ behaves in the 
near-horizon and asymptotic regimes as following:
\begin{eqnarray}
\label{real-sol}
& &{\cal R}_{n,\ell}\stackrel{x \rightarrow x_H}{\sim} 
g_{n,\ell}(x - x_H)^{\rho_n}
\left[1 + O(x -x_H)\right]   \\ \nonumber
& &{\cal R}_{n,\ell}\stackrel{x \rightarrow \infty}{\sim} 
\frac{i^{\ell + 1 + \frac{n}{2}} 2^{\frac{n}{2} - 1}}
     { \sqrt{\pi}}
\Gamma \left( \frac{1 + n}{2} \right)
Q_{n,\ell}
                                                \\  \nonumber
& & \hspace{2.0cm} \times
\left[e^{-i x + \rho_n \ln |x - x_H|} - (-1)^{\ell + \frac{n}{2}}
      S_{n,\ell}(x_H) e^{i x - \rho_n \ln |x - x_H|} \right]
+ O\left(\frac{1}{x}\right)
\end{eqnarray}
where $S_{n,\ell}(x_H)$ is a scattering amplitude and $Q_{n,\ell}$ is a 
quantity related to the multiplicities for the modes defined
\begin{eqnarray}
\label{q-factor}
Q_{n,\ell} = \left\{ \begin{array}{ll}
             \frac{\ell (\ell + n + 1) (2\ell + n + 1) 
                   (\ell + n - 1)!}{(n+2) (\ell + 1)! n!}
             &   \mbox{for vector mode photon}    \\
             \frac{(2 \ell + n + 1) (\ell + n)!}{\ell ! (n+2)!}
             &   \mbox{for scalar mode photon}    \\
             \frac{(2 \ell + n + 1) (\ell + n)!}{\ell ! (n+1)!}
             &   \mbox{for bulk scalar field}
                      \end{array}
                      \right.
\end{eqnarray}
Introducing a phase shift $\delta_{n,\ell}$ as 
$S_{n,\ell} \equiv e^{2 i \delta_{n,\ell}}$, one can 
rewrite the second equation of Eq.(\ref{real-sol}) in a form
\begin{eqnarray}
\label{real-sol2}
& &{\cal R}_{n,\ell}\stackrel{x \rightarrow \infty}{\sim} 
\frac{2^{\frac{n}{2}}}{\sqrt{\pi}} 
\Gamma \left(\frac{1 + n}{2} \right)
Q_{n,\ell}
e^{i \delta_{n,\ell}}
                                               \\  \nonumber
& & \hspace{2.0cm} \times
\sin \left[x + i \rho_n \ln |x - x_H| - \frac{\pi}{2}
           \left(\ell + \frac{n}{2}\right) + \delta_{n,\ell} \right]
+ O\left(\frac{1}{x}\right).
\end{eqnarray}
The first equation of Eq.(\ref{real-sol}) guarantees that the Wronskian
$W[{\cal R}_{n,\ell}^{\ast}, {\cal R}_{n,\ell}]_x$ is exactly same with 
Eq.(\ref{wron2}). However, Eq.(\ref{real-sol2}) implies
\begin{equation}
\label{wron4}
W[{\cal R}_{n,\ell}^{\ast}, {\cal R}_{n,\ell}]_x = -i \frac{2^n}{\pi}
\Gamma^2 \left(\frac{1 + n}{2} \right)
Q_{n,\ell}^2
e^{-2 \beta_{n,\ell}} \sinh 2 \beta_{n,\ell}
\frac{x^{n+1}}{x^{n+1} - x_H^{n+1}}
\end{equation}
where $\beta_{n,\ell} = \mbox{Im} [\delta_{n,\ell}]$. 
Thus, from Eq.(\ref{wron2}) and (\ref{wron4}) it is easy to derive a relation
\begin{equation}
\label{relat1}
|g_{n,\ell}|^2 = \frac{2^{n-2}}{\pi}
\Gamma^2 \left(\frac{1 + n}{2} \right)
Q_{n,\ell}^2
\left(1 - e^{-4 \beta_{n,\ell}} \right).
\end{equation}
Eq.(\ref{relat1}) enables us to compute the transmission coefficient 
$1 - |S_{n,\ell}|^2 = 1 - e^{-4 \beta_{n,\ell}}$ once the coefficient
$g_{n,\ell}$ is known.

Now, we would like to explain how to compute $g_{n,\ell}$ numerically. 
For the explanation 
it is convenient to introduce a new radial solution 
$\tilde{\phi}_{n,\ell}(x)$, which differs from 
${\cal R}_{n,\ell}(x, x_H)$ in its normalization in such a way that
\begin{equation}
\label{newra}
\tilde{\phi}_{n,\ell}(x, x_H)
\equiv \frac{{\cal R}_{n,\ell}}{g_{n,\ell}} \stackrel{x \rightarrow x_H}{\sim}
(x - x_H)^{\rho_n} \left[1 + O(x -x_H)\right].
\end{equation}
Since ${\cal F}_{n,\ell(\pm)}^{BL}$ in Eq.(\ref{bl-asymp1})
are two linearly independent solutions of the radial equation
Eq.(\ref{bl-radial2}), one can simply put
\begin{equation}
\label{newra2}
\tilde{\phi}_{n,\ell}(x, x_H) = f_{n,\ell}^{(-)} (x_H) 
{\cal F}_{n,\ell(+)}^{BL} (x , x_H) + f_{n,\ell}^{(+)} (x_H)
{\cal F}_{n,\ell(-)}^{BL} (x , x_H)
\end{equation}
where $f_{n,\ell}^{(\pm)}$ are called jost functions. Using Eq.(\ref{wron3}) 
one can easily compute the jost functions 
from $\tilde{\phi}_{n,\ell}$ as following
\begin{equation}
\label{jost1}
f_{n,\ell}^{(\pm)} (x_H) = \pm \frac{x^{n+1} - x_H^{n+1}}{2 i x^{n+1}} 
W[{\cal F}_{n,\ell(\pm)}^{BL}, \tilde{\phi}_{n,\ell}]_x.
\end{equation}
Inserting the explicit expressions of ${\cal F}_{n,\ell(\pm)}^{BL}$ 
presented in 
Eq.(\ref{bl-asymp1}) into Eq.(\ref{newra2}) and comparing it with the second
equation of Eq.(\ref{real-sol}), one can derive the following two 
relations
\begin{eqnarray}
\label{relat2}
S_{n,\ell}(x_H)&=& \frac{f_{n,\ell}^{(+)} (x_H)}
                        {f_{n,\ell}^{(-)} (x_H)}
                                                   \\   \nonumber
f_{n,\ell}^{(-)} (x_H)&=& \frac{2^{\frac{n}{2} - 1}}
                               {\sqrt{\pi} g_{n,\ell}(x_H)}
\Gamma \left(\frac{1 + n}{2}\right)
Q_{n,\ell}.
\end{eqnarray}
Combining Eq.(\ref{relat1}) and (\ref{relat2}) makes the greybody
factor(or transmission coefficient) of the black hole to be 
\begin{equation}
\label{jost2}
1 - |S_{n,\ell}|^2 = \frac{1}{|f_{n,\ell}^{(-)}|^2}.
\end{equation}
Thus, the partial absorption cross sections for each modes become
\begin{equation}
\label{abs1}
\sigma_{n,\ell}^{BL} = 2^{n+1} \pi^{(n+1) / 2}
\Gamma \left( \frac{3 + n}{2} \right)
Q_{n,\ell}
\frac{r_H^{n+2}}{x_H^{n+2} |f_{n,\ell}^{(-)}|^2}.
\end{equation} 
Applying the Hawking formula\cite{hawking75}, one can compute the bulk 
emission rate, {\it i.e.} the energy emitted to the bulk per unit time and
unit energy interval, as following
\begin{equation}
\label{emission1}
\frac{d^2\Gamma_{D}^{BL}}{d \omega dt} = 
\left[2^{n + 2} \pi^{(n+3) / 2} \Gamma \left( \frac{3 + n}{2} \right)
                    \right]^{-1} f_n
\frac{\omega^{n+3} \sigma_{abs}^{BL} (\omega)}
     {e^{\omega / T_H} - 1}
\end{equation}
where $\sigma_{abs}^{BL} = \sum_{\ell} \sigma_{n,\ell}^{BL}$,   
$T_H = (n+1) / 4 \pi r_H$ and
\begin{eqnarray}
\label{f-factor}
f_n = \left\{ \begin{array}{ll}
              n + 2       &   \mbox{for photon modes}   \\
              1           &   \mbox{for bulk scalar field}
              \end{array}
              \right.
\end{eqnarray}

The jost functions $f_{n,\ell}^{(\pm)}(x_H)$ can be computed numerically
by adopting the analytic continuation. In order to apply the continuation
we need a solution of Eq.(\ref{bl-radial2}), which is convergent in the 
neighborhood of $x=b$, where $b$ is an arbitrary point. Thus the expression
of the solution is 
\begin{equation}
\label{bl-int1}
\tilde{R}_{n,\ell} (x) = (x - x_H)^{\rho_n} 
\sum_{N=0}^{\infty} D_N (x - b)^N.
\end{equation}
The recursion relation for the coefficients $D_N$ can be 
explicitly derived by inserting 
Eq.(\ref{bl-int1}) into (\ref{bl-radial2}). Since it is too lengthy, we will not
present the explicit expression. Using a solution (\ref{bl-int1}), 
one can increase the 
convergent region for the near-horizon solution from the near-horizon regime 
and decrease the convergent region for the asymptotic solution from the 
asymptotic regime. Repeating the procedure eventually makes the two 
solutions which have common convergent region. Then one can compute the jost
functions by making use of these two solutions and Eq.(\ref{jost1}).

Fig. 1 is a plot of the emission spectrum $d^2 \Gamma / d \omega d t$ for the
spin-$1$ photon when there is no extra dimension. For a comparision the
spectrum for the spin-$0$ scalar field is plotted together. Fig. 1 shows
that the emission rate for the scalar field is much larger than that for the
photon field when $n = 0$. 

However, the situation is drastically changed when
the extra dimensions exist. In Fig. 2 the emission spectra for the scalar
photon mode, vector photon mode, and bulk scalar field are plotted
when $n = 1$ (Fig. 2(a)), $n = 2$ (Fig. 2(b)), $n = 4$ (Fig. 2 (c)) and
$n = 6$ (Fig. 2(d)). Fig. 2 indicates that the emission rate in general
increases with increasing $n$ regardless of the type of fields. However,
the increasing rates of the emission spectra for the photon modes are
much larger than that for the bulk scalar field. 

Table I shows the total emission rate, {\it i.e.}
$\int d \omega d^2 \Gamma / d \omega dt$ for the scalar mode, vector mode
and bulk scalar when $n = 0$, $1$, $2$, $4$ and $6$. The table indicates 
that the total emissivity for the photon field is only $23 \%$ of that
for the spin-$0$ field when there is no extra dimension. However, this 
ratio factor increases rapidly with increasing $n$. For example, this 
factor becomes $172 \%$, $321 \%$, $499 \%$ and $790 \%$ when 
$n = 1$, $2$, $4$ and $6$. 

Another interesting feature is that for comparatively small $n$ the emission
rate for the scalar mode photon is larger than that for the vector mode photon.
However, the increasing rate for the vector mode is much larger compared
to that for the scalar mode with increasing $n$. As a result, the total 
emissivity for the vector mode becomes dominant more and more in the 
photon emission spectrum for large $n$. For example, the emissivity for the
vector mode is roughly three times than that for the scalar mode when
$n = 6$.

\begin{figure}[ht!]
\begin{center}
\includegraphics[height=10cm]{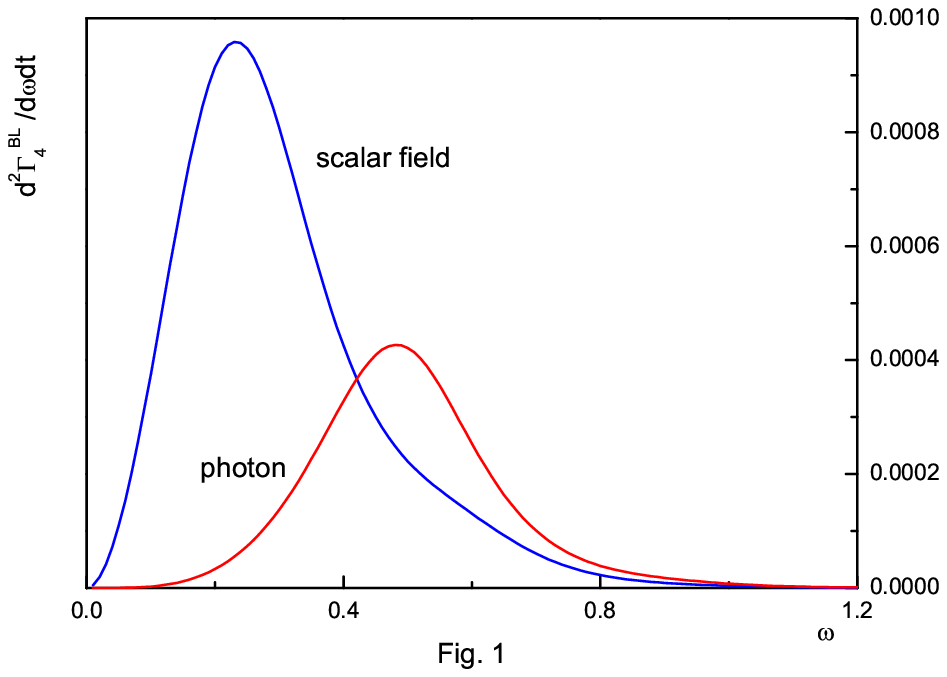}
\caption[fig1]{Plot of the emission spectra for the spin-$1$ photon and spin-$0$
scalar fields when there is no extra dimension. This figure indicates that the 
emission rate for the photon is much smaller than that for the scalar
when $n=0$.}
\end{center}
\end{figure}

\begin{figure}[ht!]
\begin{center}
\includegraphics[height=6.3cm]{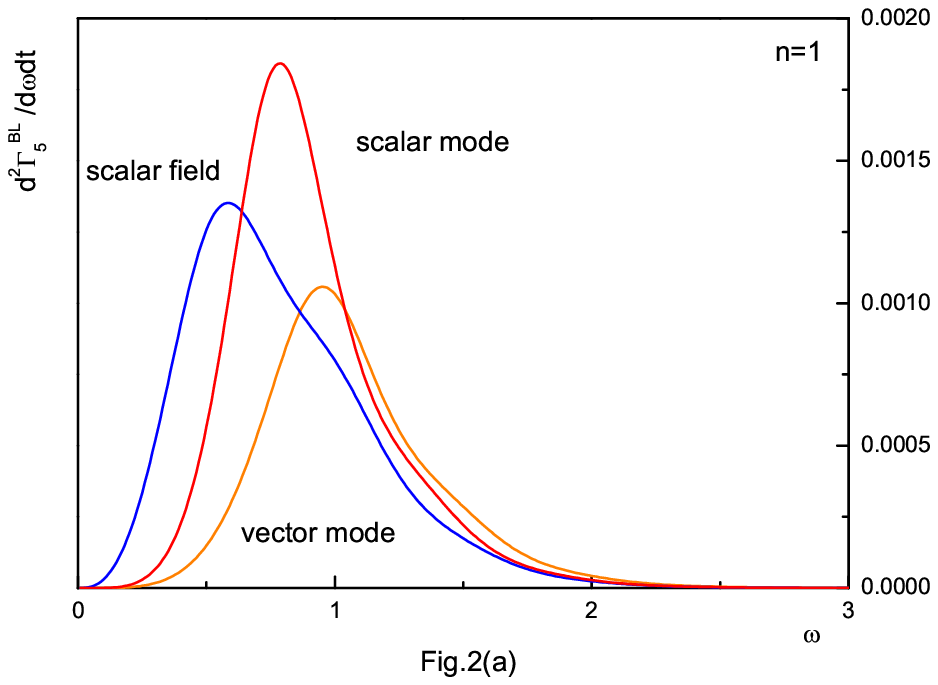}
\includegraphics[height=6.3cm]{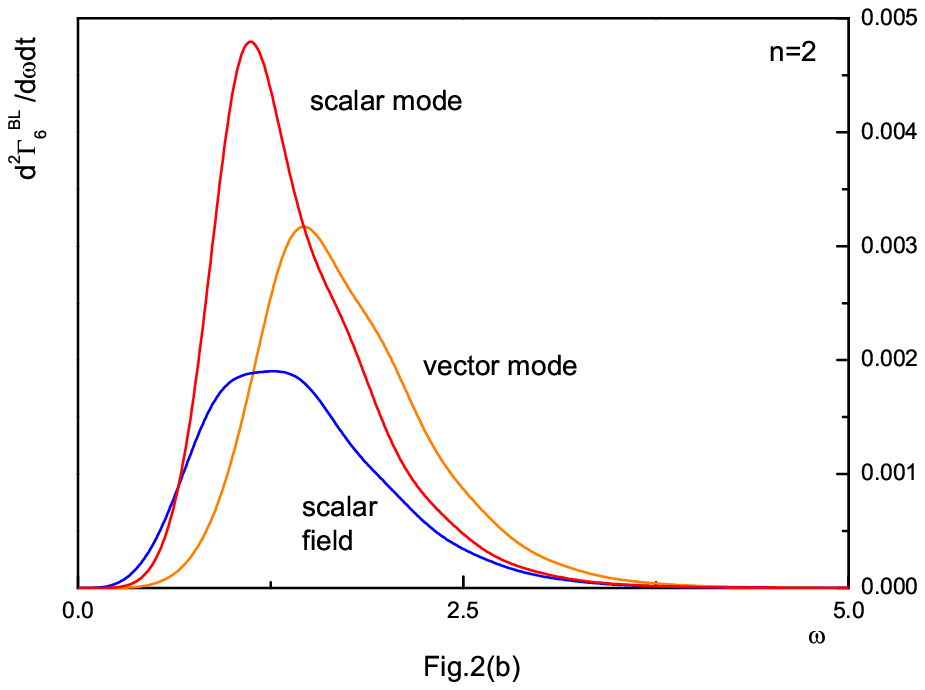}
\includegraphics[height=6.3cm]{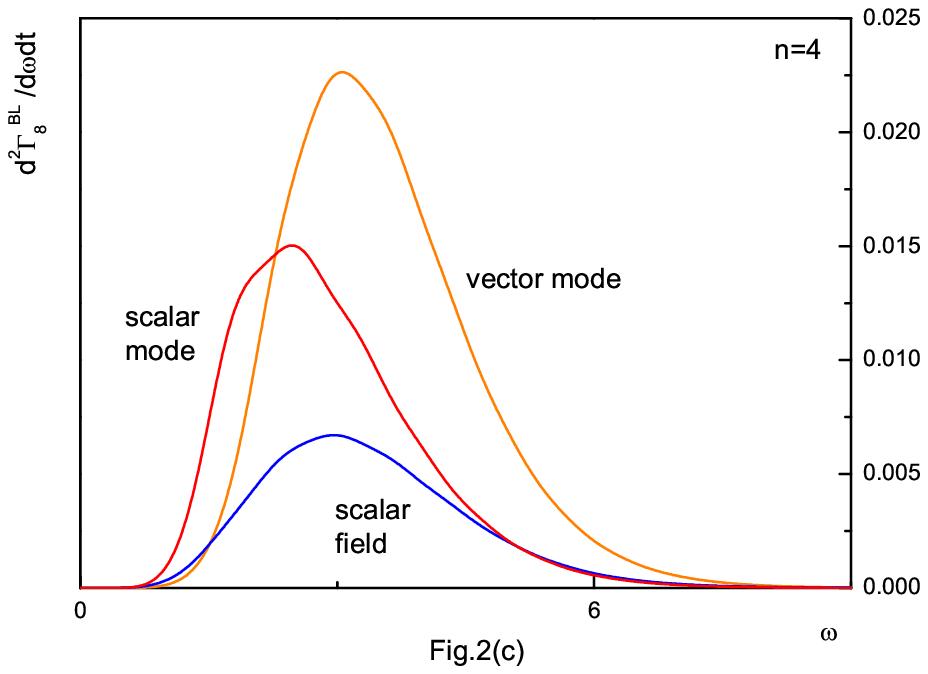}
\includegraphics[height=6.3cm]{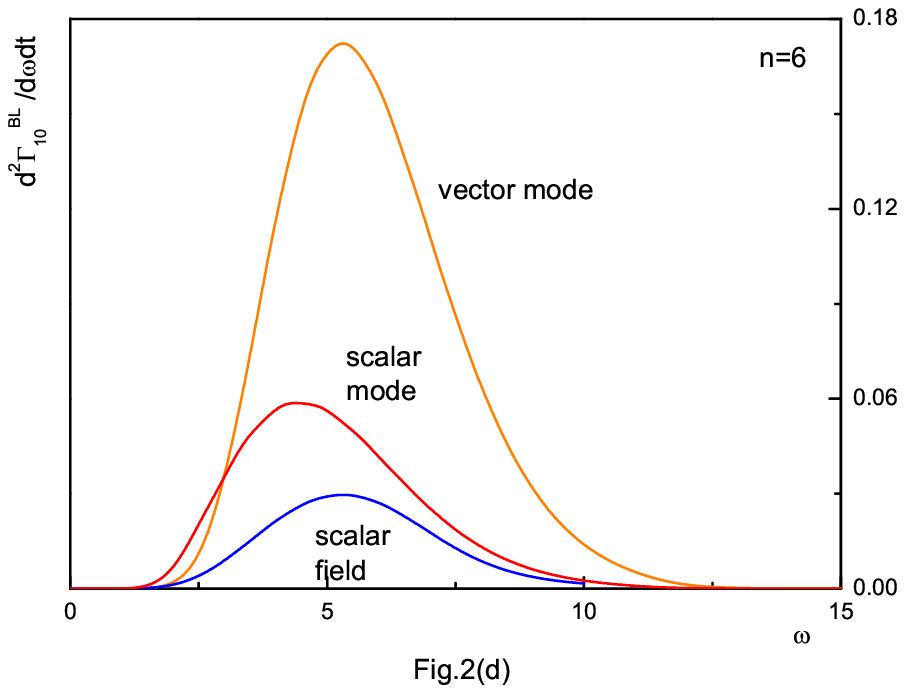}
\caption[fig2]{Plot of the emission spectra for the bulk photon modes when
$n=1$ (a), $n=2$ (b), $n=4$ (c) and $n=6$ (d). The spectra for the bulk scalar
fields are plotted together for a comparision. The figures indicate that the 
emission rates in general increase with increasing $n$ regardless of the type
of fields. However, they also indicate that as $n$ increases, the emission rates
for the photon fields become more and more dominant compared to those for the 
bulk scalar fields. }
\end{center}
\end{figure}

\begin{center}

\begin{tabular}{l|l|l|l|l|r} \hline
$ $           &   $4d$   &     $5d$     &     $6d$   &    $8d$  
                                              &   $10d$  \\  \hline \hline
scalar mode photon   &   $6.728 \times 10^{-5}$   &  $1.088 \times 10^{-3}$  &  
$4.77 \times 10^{-3}$ &  $3.64 \times 10^{-2}$  &  $0.2451$     \\
vector mode photon  &   $6.728 \times 10^{-5}$  &  $7.21 \times 10^{-4}$    &  
$3.82 \times 10^{-3}$  &  $5.73 \times 10^{-2}$  &  $0.708$    \\ 
bulk scalar field  &  $2.975 \times 10^{-4}$   &  $1.05 \times 10^{-3}$   &   
$2.68 \times 10^{-3}$  &  $1.876 \times 10^{-2}$  &  $0.1206$        \\ 
                                                              \hline \hline 
photon / bulk scalar &  $0.46$  &  $1.72$  &  $3.21$  &  $4.99$  &  $7.90$  \\
\hline
\end{tabular}

\vspace{0.1cm}
Table I: Relative Emission Rates for Bulk Photon and Scalar Fields 
\end{center}
\vspace{1.0cm}

\section{Brane Photon}
In this section we would like to compute the emissivity of the spin-$1$
photon on the brane by the induced metric 
\begin{equation}
\label{induce1}
ds_4^2 = -h dt^2 + h^{-1} dr^2 + r^2 (d\theta^2 + \sin^2 \theta d \phi^2).
\end{equation}
In fact, this was already computed in Ref.\cite{kanti1} by different 
numerical method. We would like to re-calculate the emission rates of the
brane-localized photon field by
incorporating our numerical method into the complex potential method 
introduced in Ref.\cite{chandra92} 
to compare the results with those for the bulk photon.

The radial equation for the scalar ($s = 0$), fermion ($s = 1/2$), and 
vector ($s = 1$) can be expressed as a following master equation\cite{ida02} :
\begin{equation}
\label{master}
\Lambda^2 Y + P \Lambda_- Y - Q Y = 0
\end{equation}
where $\Lambda_{\pm} = d / d r_* \pm i \omega$, 
$\Lambda^2 \equiv \Lambda_+ \Lambda_-$, $d / d r_* \equiv h d / dr$ and
\begin{eqnarray}
\label{PandQ}
P&=&\frac{d}{d r_*} \ln \left(\frac{r^2}{h} \right)^{-s}
                                                        \\  \nonumber
Q&=&\frac{h}{r^2} \left[{\cal A}_{\ell s} + (2 s + n + 1) (n s + s + 1)
                        (1 - h) \right]
\end{eqnarray}
with ${\cal A}_{\ell s} = \ell (\ell + 1) - s (s + 1)$.
Defining $Y = f R$ with $(1/f) d f / d r_* = - P / 2$. one can change
Eq.(\ref{master}) into the Schr\"{o}dinger-like expression with a complex
potential in the following:
\begin{equation}
\label{br-radial1}
\Lambda^2 R = V_s R
\end{equation}
where
\begin{eqnarray}
\label{br-poten1}
V_s&=& i \omega P + \frac{P^2}{4} + \frac{1}{2} \frac{d P}{d r_*} + Q
                                                            \\  \nonumber
&=& \frac{h}{r^2}
\left[ {\cal A}_{\ell s} + q_{n s} (1 - h) + \frac{s^2}{4 h}
      \left\{ (n + 1) (1 - h) - 2 h \right\}^2 + s h \right]
+ \frac{i \omega s}{r} \left\{ (n + 1) (1 - h) - 2 h \right\}
\end{eqnarray}
and $q_{n s} = (2 s + n + 1) (n s + s + 1) - (s/2) (n + 1) (n + 4)$.

Defining the dimensionless parameters $x = \omega r$ and $x = \omega r_H$, 
we can rewrite Eq.(\ref{br-radial1}) in the following form:
\begin{eqnarray}
\label{br-radial2}
& &x^2 \left(x^{n+1} - x_H^{n+1} \right)^2 \frac{d^2 R}{d x^2} + 
(n+1) x_H^{n+1} x \left(x^{n+1} - x_H^{n+1} \right) \frac{d R} {d x}
                                                        \\  \nonumber
& & - \Bigg[ i s x^{n+2} \left\{(n+3) x_H^{n+1} - 2 x^{n+1} \right\}
             + {\cal A}_{\ell s} x^{n+1} \left(x^{n+1} - x_H^{n+1} \right)
             + q_{n s} x_H^{n+1}  \left(x^{n+1} - x_H^{n+1} \right) 
                                                         \\  \nonumber
& & \hspace{3.0cm} 
             + \frac{s^2}{4} \left\{(n+3) x_H^{n+1} - 2 x^{n+1} \right\}^2
             - x^{2 n + 4} + s \left(x^{n+1} - x_H^{n+1} \right)^2 \Bigg]
R = 0.
\end{eqnarray}
Using Eq.(\ref{br-radial2}) one can easily show that if $R_1$ and $R_2$ are 
solutions of Eq.(\ref{br-radial2}), the Wronskian between them reduces to 
\begin{equation}
\label{wron11}
W[R_1, R_2]_x \equiv R_1 \frac{d R_2}{d x} - R_2 \frac{d R_1}{d x}
= {\cal C} \frac{x^{n+1}}{x^{n+1} - x_H^{n+1}}
\end{equation}
where ${\cal C}$ is an integration constant.

The solution of Eq.(\ref{br-radial2}) which is convergent in the near-horizon
regime can be written as 
\begin{equation}
\label{br-near1}
{\cal G}_{n,\ell}^{BR} (x, x_H) = \omega^{s - 1 + i x_H}
x_H^{\frac{s}{2} (n - 1) + 1}
\left[(n+1) x_H^n\right]^{-\frac{s}{2} - i \frac{x_H}{n+1}}
\sum_{N=0}^{\infty} d_{\ell,N} (x - x_H)^{N + \rho_n}
\end{equation}
with $d_{\ell,0} = 1$ and 
\begin{equation}
\label{multi1}
\rho_n = \pm \left[\frac{s^2}{4} - \frac{x_H^2}{(n + 1)^2} + i s 
                  \frac{x_H}{n + 1} \right]^{1/2}.
\end{equation}
The sign of $\rho_n$ is chosen by $\mbox{Im} \rho_n < 0$. The 
multiplication constant in Eq.(\ref{br-near1}) is chosen for the later 
convenience. The recursion relation for the coefficient $d_{\ell,n}$ can be
derived explicitly by inserting Eq.(\ref{br-near1}) into (\ref{br-radial2}).

The ingoing and outgoing solutions which are convergent in the asymptotic 
regime are respectively
\begin{eqnarray}
\label{br-asymp1}
{\cal F}_{n,\ell (+)}^{BR} (x, x_H)&=& \omega^{-s} x^{s + 1} e^{-i x}
                     \sum_{N = 0}^{\infty} \tau_{N (+)} x^{-(N + 1)}
                                                    \\   \nonumber
{\cal F}_{n,\ell (-)}^{BR} (x, x_H)&=& \omega^{s} x^{-s + 1} e^{i x}
                     \sum_{N = 0}^{\infty} \tau_{N (-)} x^{-(N + 1)}.
\end{eqnarray}
Of course, the recursion relations for $\tau_{N (\pm)}$ can be explicitly 
derived by making use of the radial equation (\ref{br-radial2}). 
Eq.(\ref{wron11}) guarantees that the Wronskian between 
${\cal F}_{n,\ell (+)}^{BR}$ and ${\cal F}_{n,\ell (-)}^{BR}$ is 
\begin{equation}
\label{wron12}
W[{\cal F}_{n,\ell (+)}^{BR}, {\cal F}_{n,\ell (-)}^{BR}]_x = 
2 i \frac{x^{n + 1}}{x^{n + 1} - x_H^{n + 1}}.
\end{equation}

In order to compute the absorption and emission spectra we define the 
near-horizon and asymptotic solutions in the following:
\begin{eqnarray}
\label{br-two}
R_{NH}(x)&=&A_n {\cal G}_{n,\ell}^{BR} (x, x_H)
                                             \\   \nonumber
R_{\infty}(x)&=& {\cal F}_{n,\ell (+)}^{BR}(x, x_H) + 
B_n {\cal F}_{n,\ell (-)}^{BR}(x, x_H).
\end{eqnarray}
Using (\ref{wron12}), one can compute the coefficients $A_n$ and $B_n$
\begin{eqnarray}
\label{br-coef}
A_n&=&- \frac{2 i x^{n+1}}{x^{n+1} - x_H^{n+1}}
\frac{1}{W[{\cal F}_{n,\ell (-)}^{BR}, {\cal G}_{n,\ell}^{BR}]_x}
                                                 \\  \nonumber
B_n&=& - \frac{W[{\cal F}_{n,\ell (+)}^{BR}, {\cal G}_{n,\ell}^{BR}]_x}
              {W[{\cal F}_{n,\ell (-)}^{BR}, {\cal G}_{n,\ell}^{BR}]_x}.
\end{eqnarray}

The method of the complex potential introduced in Ref.\cite{chandra92} 
makes that the
transmission coefficient reduces to $|A_n|^2 / r_H^2$. Thus 
the absorption cross section for the photon field becomes
\begin{equation}
\label{br-abs1}
\sigma_{n,\ell}^{BR} = \frac{\pi (2 \ell + 1)}{(\omega r_H)^2}
|A_n|^2.
\end{equation}
Applying the Hawking formula, one can compute the brane emission rate 
$d^2 \Gamma_D^{BR} / d \omega d t$, {\it i.e.} the energy emitted on the 
brane per unit time and unit energy interval:
\begin{equation}
\label{br-emi1}
\frac{d^2 \Gamma_D^{BR}}{d \omega d t} = \tilde{f}_s (2 \pi^2)^{-1}
\frac{\omega^3 \sigma_{abs}^{BR} (\omega) } {e^{\omega / T_H} - 1}
\end{equation}
where $\sigma_{abs}^{BR} = \sum_{\ell} \sigma_{n,\ell}^{BR}$, 
$T_H = (n+1) / 4 \pi r_H$ and $\tilde{f}_s = 2$ (or $1$) for photon
field (or brane scalar field).

\begin{figure}[ht!]
\begin{center}
\includegraphics[height=10cm]{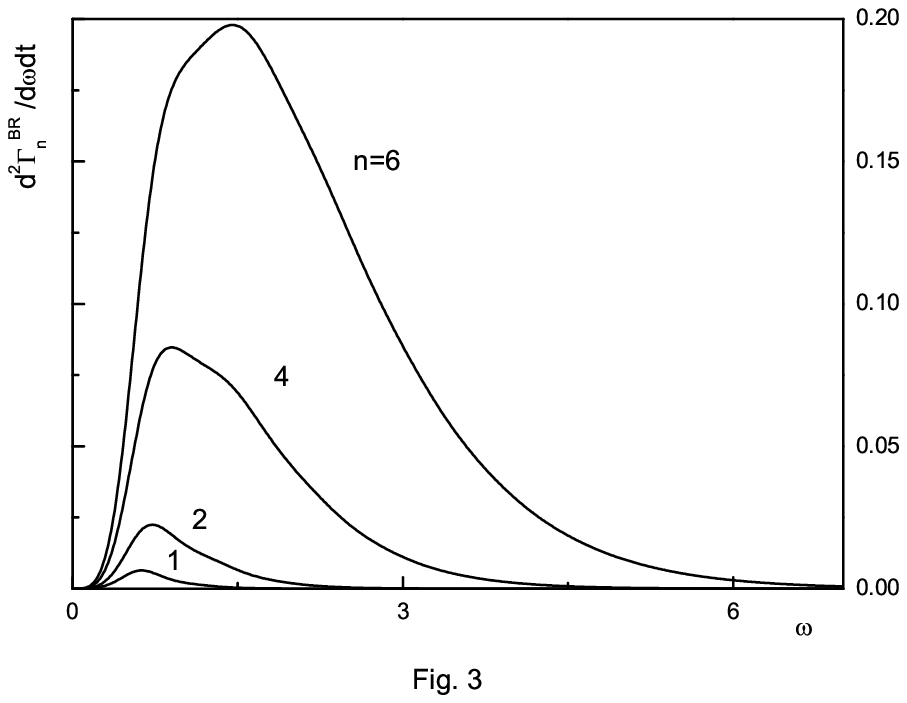}
\caption[fig3]{The $n$-dependence of the emission spectra for the brane-localized
spin-$1$ photon fields. With increasing $n$ the emission rates tend to increase
rapidly.   }
\end{center}
\end{figure}

In Fig. 3 the $n$-dependence of the emission spectra on the brane 
for the spin-$1$ photon fields is 
plotted when $n=1$, $2$, $4$, and $6$. As it is well-known, the existence 
of the extra dimensions in general enhances the emissivity. Fig. 3 indicates 
that the increasing
rate of the emissivity is very rapid with increasing $n$. In the next section
we will compare the bulk and brane emissivities for the spin-$1$ photon
fields.

\section{Bulk versus Brane}
In Ref.\cite{emp00} EHM argued that the higher-dimensional black holes mainly 
radiate on the brane. This argument was confirmed in Ref.\cite{kanti1}, 
where the 
Hawking emissivities of the bulk and brane-localized scalar fields were
numerically compared  in the background of the $(4 + n)$-dimensional 
Schwarzschild black hole. If one defines the total emission rates 
\begin{equation}
\label{tot-emi}
\tilde{\Gamma}_D^{BL} \equiv \int d \omega
\frac{d^2 \Gamma_D^{BL}}{dt d\omega}
\hspace{2.0cm}
\tilde{\Gamma}_D^{BR} \equiv \int d \omega
\frac{d^2 \Gamma_D^{BR}}{dt d\omega},
\end{equation}
the ratio $\tilde{\Gamma}_D^{BL} / \tilde{\Gamma}_D^{BR}$ in the scalar
emission is summarized in Table II. The Table II indicates that the 
emission on the brane is dominant when $D \leq 11$ for the case of scalar
fields.

\begin{center}

\begin{tabular}{l|l|l|l|l|l|l|l|r} \hline   \hline
$ $           & $D=4$ & $D=5$ & $D=6$ & $D=7$
         &  $D=8$  &  $D=9$  &  $D=10$  &  $D=11$ \\  \hline 
$\tilde{\Gamma}_D^{BL} / \tilde{\Gamma}_D^{BR}$   &   $1.0$ &  $0.40$  &
$0.24$ &  $0.22$  &  $0.24$  &  $0.33$  &  $0.52$  &  $0.93$     \\
\hline
\end{tabular}

\vspace{0.1cm}
Table II: Relative Bulk-to-Brane Emissition Rates for Scalar Fields
\end{center}

Now, we would like to summarize the result of the spin-$1$ photon emission
rate in the background of the $(4+n)$-dimensional Schwarzschild phase.
In Fig. 4 we plot the emission spectra for the brane-localized and bulk 
photons together for a comparision when $n=1$ (Fig. 4(a)), $n=2$ (Fig. 4(b)),
$n=4$ (Fig. 4(c)) and $n=6$ (Fig. 4(d)). Fig. 4 indicates that the emission 
on the brane is dominant when $n$ is not too large. However, the emission rates
for the bulk vector mode photon rapidly increase with increasing $n$. 
Therefore, 
the peak point of the emission spectrum for the vector mode becomes roughly 
same with that for the brane-localized photon when $n=6$. 

For more precise comparision the relative bulk-to-brane emission rates are 
summarized in Table III. Table III indicates that the brane emission is a 
bit dominant when $D \leq 8$. But the dominance is changed into the total bulk 
emission when $D \geq 10$. Since the bulk photon has $n+2$ polarization states
while brane photon has $2$ helicities, it is also possible to compute the 
bulk-to-brane emissivity per degree of freedom(d.o.f.) which is shown in
Table III. In spite of the rapid increase of the emission rates for the 
bulk photon with increasing $n$, this ratio always remains smaller than 
unity, which strongly supports the EHM argument.

Comparing Table III with Table II, one can realize that the effect of nonzero
spin tends to make the bulk emission to be dominant more easily. Thus it seems 
to be very interesting to examine the graviton emission problem, where the 
effect of the spin may be more strong.

\begin{figure}[ht!]
\begin{center}
\includegraphics[height=6.3cm]{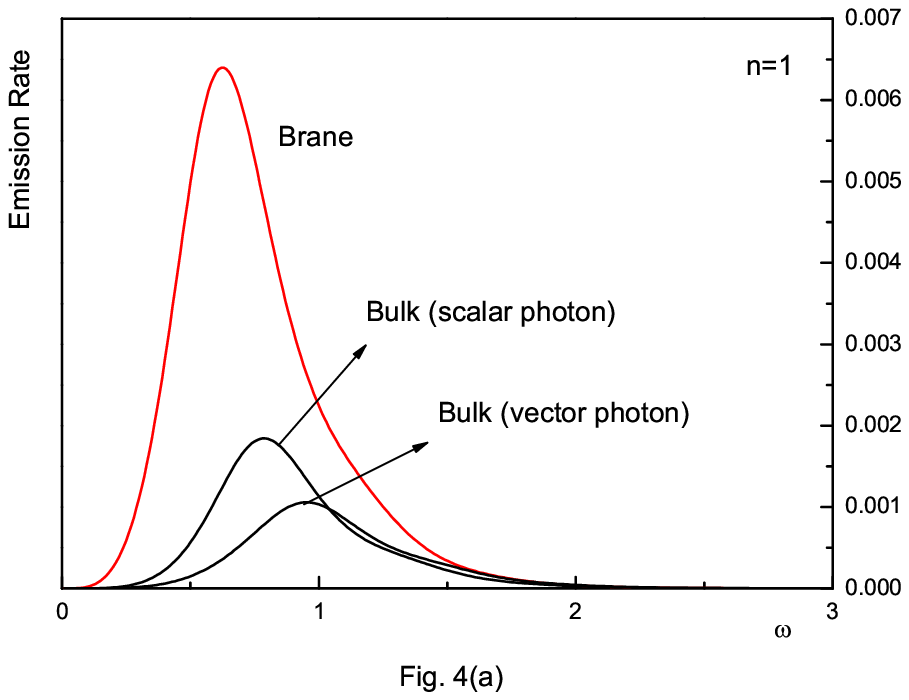}
\includegraphics[height=6.3cm]{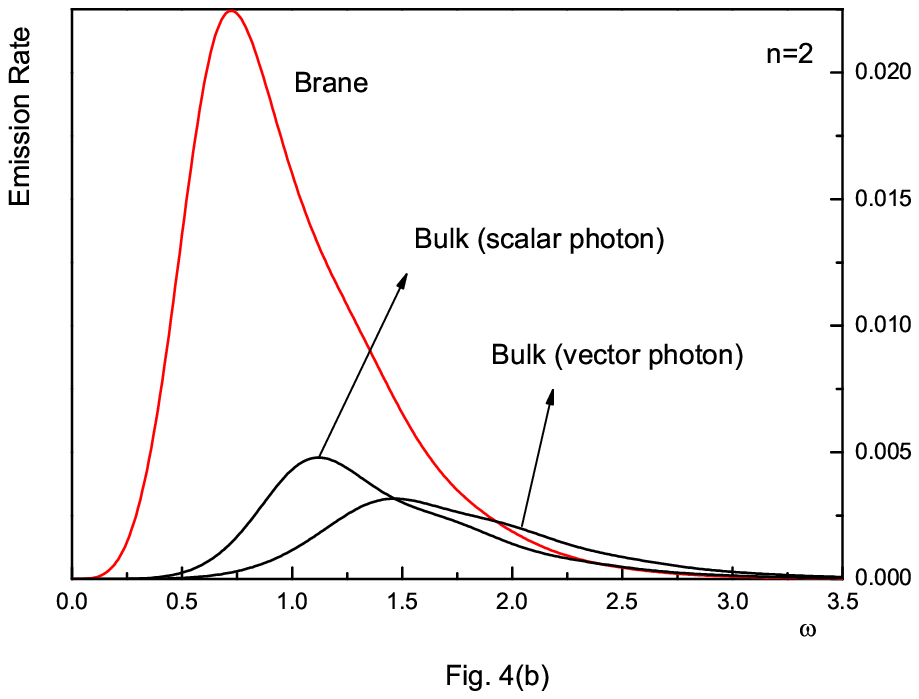}
\includegraphics[height=6.3cm]{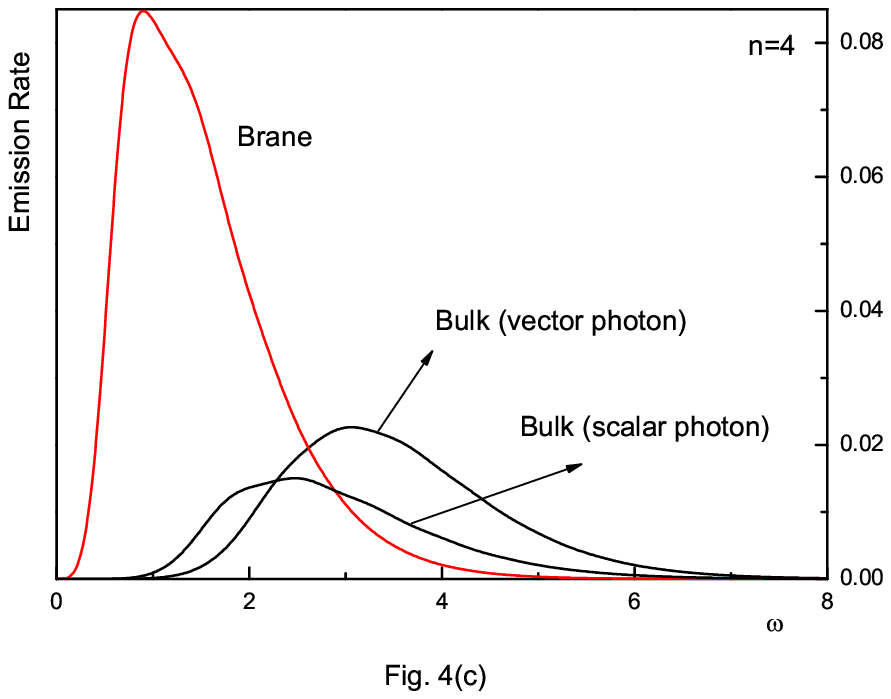}
\includegraphics[height=6.3cm]{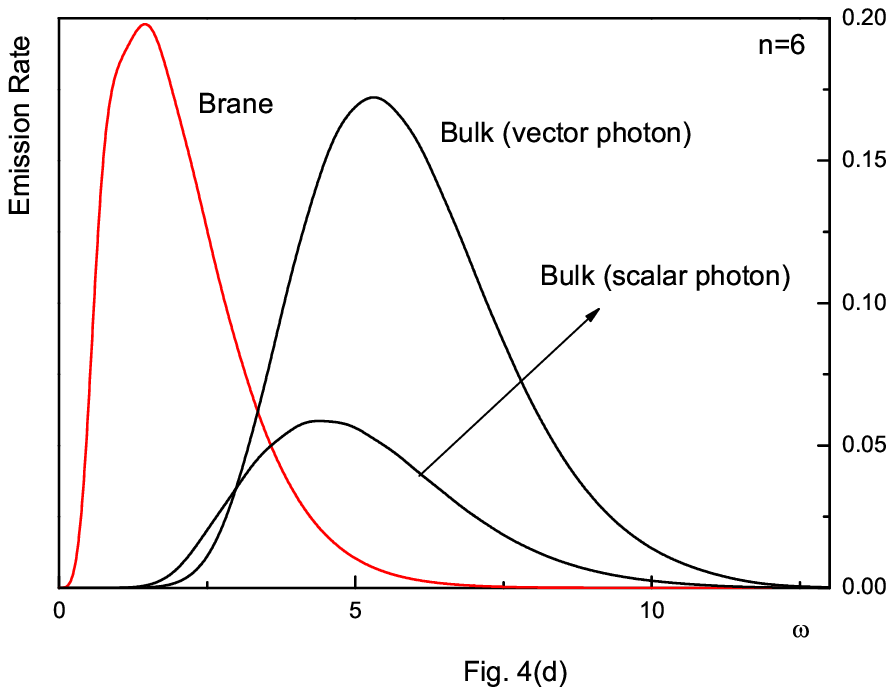}
\caption[fig4]{Plot of the emission spectra for the brane-localized and bulk 
photons when $n=1$ (a), $n=2$ (b), $n=4$ (c) and $n=6$ (d). The figures indicate 
that the emission on the brane is a bit dominant when $n$ is not too large. 
However, the emission rates for the bulk vector mode photon increase very rapidly
with increasing $n$ compared to other fields. Thus the dominance of the 
brane emission is no longer true when $D \geq 10$.       }
\end{center}
\end{figure}

\begin{center}

\begin{tabular}{l|l|l|l|l|l|r} \hline   \hline
$ $           & $D=4$ & $D=5$ & $D=6$ & $D=7$ &
          $D=8$  &    $D=10$  \\  \hline
Scalar Mode / Brane   &   $0.50$ &  $0.30$  &
$0.25$ &  $0.24$  &  $0.26$  &  $0.51$       \\
\hline
Vector Mode / Brane  &  $0.50$  &  $0.20$ & $0.20$ & $0.26$ & $0.42$  &  $1.49$
                                                                \\   \hline
Bulk / Brane & $1.0$ & $0.50$ & $0.44$ & $0.50$ &  $0.68$  &  $2.0$ \\  \hline
(Bulk / Brane) per d.o.f. & $1.0$ & $0.33$ & $0.22$ & $0.20$ & $0.23$ & $0.50$
                                                                 \\    \hline
\end{tabular}

\vspace{0.1cm}
Table III: Relative Bulk-to-Brane Emissition Rates for Photon Fields
\end{center}

\section{Conclusion}
In this paper we computed the absorption and emission spectra for the 
bulk and brane-localized spin-$1$ photon fields when the spacetime 
background is a $(4+n)$-dimensional Schwarzschild phase. For the case of
the bulk photon field we compute the spectra for the scalar mode and 
vector mode separately. The total emissivity for the scalar mode photon
is dominant compared to that for the vector mode photon for comparatively
small $n$. However, the increasing rate of the emission rate for the vector
mode is much larger than that for the scalar mode with increasing $n$. As
a result, the total emissivity for the vector mode photon becomes dominant
more and more in the bulk emission rate for large $n$. 
For example, the vector-to-scalar relative emissivities become 
$66.3\%$, $80.1\%$, $157.4\%$ and $288.9\%$ when $n=1$, $2$, $4$ and $6$
respectively. 

The absorption and emission spectra for the brane-localized photon field are
also computed using the complex potential method introduced in 
Ref.\cite{chandra92}. The total emissivity for the brane-localized spin-$1$
field tends to increase with increasing $n$ like the case of the bulk photon
field.  

Comparing the total emissivity for the bulk photon to that for the 
brane-localized photon, we can see that the latter is a little bit larger than 
the former when $n$ is small like a scalar field. Of course, the former 
becomes dominant more and more with increasing $n$. Comparision of 
Table III with Table II indicates that the effect of the non-zero spin
tends to make the bulk emission to be dominant more easily. The dominance
for the bulk photon seems to be mainly due to the fact that the bulk
photon has larger d.o.f. than the brane photon. Thus we compute the 
bulk-to-brane relative emissivity per d.o.f., which is shown in Table III.
In spite of rapid increase of the bulk emissivity, this ratio always remains
smaller than unity, which supports the EHM argument. However, the rapid
increase of the total bulk emissivity indicates that the missing energy
in the future collider experiment is not negligible compared to the visible
one. Thus, one should carefully consider the portion of the missing energy
when the experiment in the future colliders is setting up. In order to 
explore the missing energy more exactly 
it is important to examine the graviton emission problem.

The emission rates for the bulk graviton were discussed in detail in 
Ref.\cite{corn95}. However, the rates for the brane graviton are not
computed completely. In fact, the emission spectra for the axial mode
graviton were computed in Ref.\cite{park06-1}. Since the 
induced metric (\ref{induced}) is not a vacuum solution of $4d$ Einstein
field equation, the author in Ref.\cite{park06-1} tried to determine the
components of the energy-momentum tensor by making use of the general 
principles. However, the energy-momentum tensor is not fully determined
from the general principles and therefore an assumption was used to 
fix the tensor. 

To compute the emission rates for the brane-localized graviton without 
invoking the assumption, we think we should rely on the complex potential
method again. For this case there seem to be two non-trivial tasks. Firstly,
we should check whether Eq.(\ref{master}) is a really master equation including
the spin-$2$ graviton fields. As Ref.\cite{kanti1} has shown, Eq.(\ref{master})
holds for the scalar, fermion and electromagnetic fields. However, it is 
not shown explicitly that Eq.(\ref{master}) holds for the spin-$2$ graviton
field\footnote{However, Ref.\cite{berti03} used Eq.(\ref{master}) for the 
computation of  the quasinormal
modes of the gravitational perturbation.}. Secondly, in Ref.\cite{chandra92} 
the greybody factor for the graviton field was computed by making use of 
the Hawking-Hartle theorem\cite{hawk72} when there is no extra dimension. 
It seems to be formidable job to extend the theorem when there are extra 
dimensions because the induced metric is not a vaccum solution of the 
$4d$ Einstein equation. We hope to discuss this issue in the near future.

\vspace{1cm}

{\bf Acknowledgement}:  
This work was supported by the Kyungnam University
Research Fund, 2006.

\end{document}